# Analysis of the bandwidth of resonant spectra for tilted fiber grating

**Liang Fang, Hongzhi Jia***

*Engineering Research Center of Optical Instrument and System, Ministry of Education, Shanghai Key Lab of Modern Optical System, University of Shanghai for Science and Technology, No. 516 JunGong Road, Shanghai 200093, China*

**Corresponding author: hzjia@usst.edu.cn*

**Abstract:**

Based on coupled-mode equations, the formula of null-null bandwidth for tilted fiber grating is deduced in this paper. Numerical simulations and theoretical analysis of the effects of the tilt angle on bandwidth and maximum reflectivity are performed. Furthermore, under the maximum reflection of 95%, the relation of bandwidth with modulation amplitude and length of grating, and the needed length of grating are demonstrated in two three-dimensional diagrams. It reveals that the grating tilt can obviously narrow the bandwidth, especially in the case of large modulation amplitude and grating length, which is meaningful to its application on optical communication and fiber sensing.

## 1. Introduction

Optical fiber gratings have been developed intensively in recent decades, and have been applied extensively in many areas. Up to present, there has been a wide variety of fiber gratings being reported. The typical types of gratings include uniform, apodized, chirped, phase-shifted, tilted and long-period gratings, each of which has its particular spectral characteristics and applications, especially in optical communication and fiber sensing [1-3]. For tilted gratings, in the core-mode reflection, it is well known that grating tilt reduces the efficiency of coupling between forward and backward core-modes, and hence reduces the reflectivity [1, 4, 5]. However, it has not been paid much attention that grating tilt has great influence on the bandwidth of reflective spectrum. Therefore, this article focuses on theoretical analysis of the effect of grating tilt on the bandwidth.

The bandwidth of fiber Bragg gratings (FBGs) has been analyzed in many papers, it is the modulation amplitude and length of FBG that influence the bandwidth [1]. In tilted gratings, the grating tilt is also a factor to act on bandwidth. Therefore, in this article, we deduce the formulas of reflectivity and bandwidth with tilt angle based on coupled-mode equations. In addition to the theoretical formalism, the simulated reflective spectra are performed under different tilt angle. In order to make clear the comprehensive effects of grating length and modulation amplitude on bandwidth, two three-dimensional diagrams are plotted in the case of the maximum reflection of 95%. There are some special regions corresponding to grating length and modulation, where the bandwidth of reflective spectrum is extraordinary narrow, which is practically meaningful to the tilted grating's application on optical communication and fiber sensing.

## 2. Theory treatment

For uniform tilted grating, a perturbation on the effective refractive index $n_{eff}$ of the guided modes can be simply described by

$$\delta n_{eff}(z') = \overline{\overline{\delta n_{eff}}}(z')\left[1+\cos\left(\frac{2\pi}{\Lambda_g} z'\right)\right] \tag{1}$$

where $\Lambda_g$ is the grating period along the $z'$-axis, shown Fig. 1, $\overline{\overline{\delta n_{eff}}}(z')$ is the "dc" index change spatially averaged over a grating period. In the slowly varying functions, we take approximately $z' = z\cos\theta + x\sin\theta \approx z\cos\theta$.

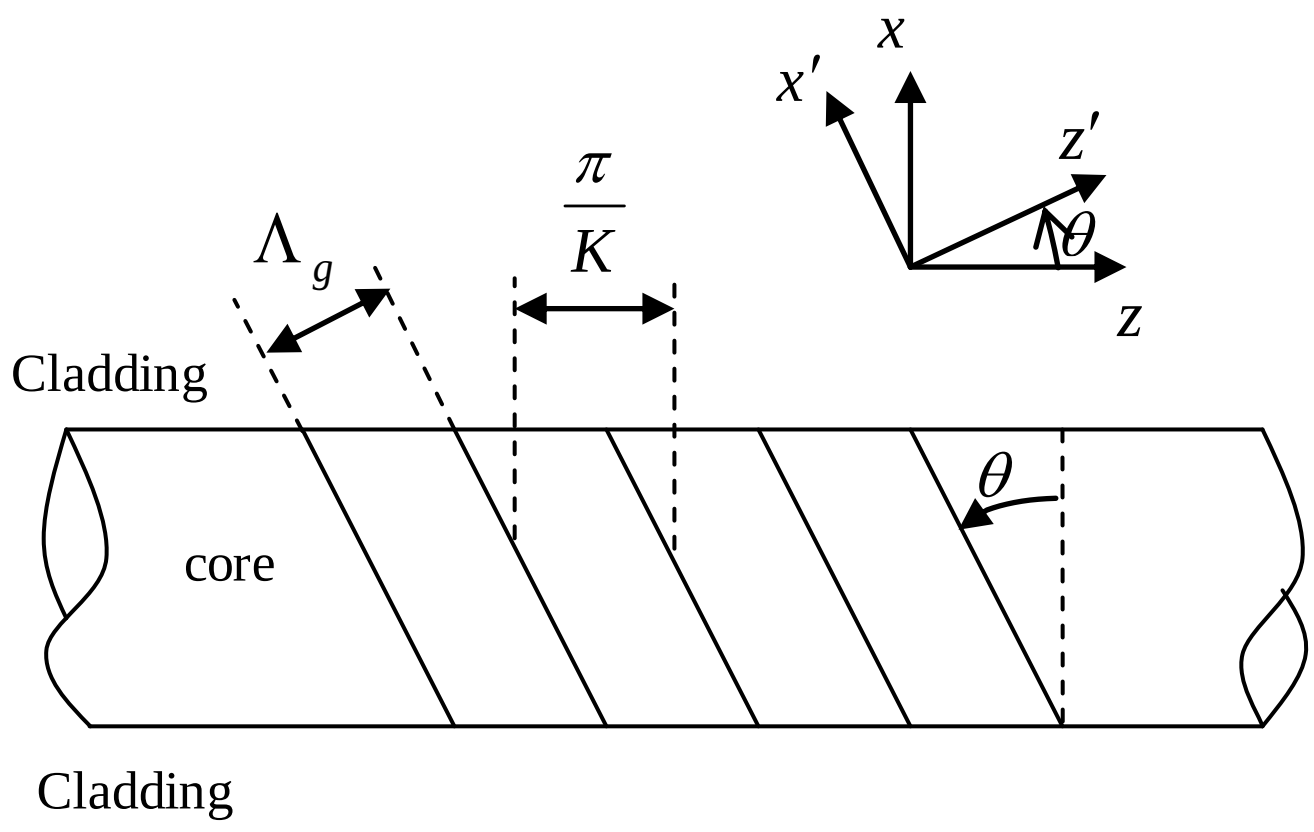

Fig. 1. Configuration of tilted fiber grating.

According to the reference [4], coupled-amplitude equations can be described as follows:

$$\begin{aligned} \frac{dR}{dz} &= i\left[\bar{\beta} g_f(\theta)\sigma(z) + \beta_{01} - K\right]R(z) + i\bar{\beta} g_b(\theta)\kappa(z)S(z) \\ \frac{dS}{dz} &= -i\left[\bar{\beta} g_f(\theta)\sigma(z) + \beta_{01} - K\right]S(z) - i\bar{\beta} g_b(\theta)\kappa(z)R(z) \end{aligned} \tag{2}$$

Where $R(z)$ and $S(z)$ are forward-propagating and backward-propagating core-mode amplitude related variables in tilted fiber gratings, respectively. To simplify the notation, approximately, $\kappa(z) \equiv \bar{\kappa}(z')$, describes the grating index modulation amplitude, and $\sigma(z) \equiv \bar{\sigma}(z')$, describes the concomitant slowly varying perturbation in the background index of refraction that accompanies the tilted grating writing[4,6], which are constants for uniform gratings , here

$$\sigma(z) = 2\kappa(z) \approx \overline{\delta n_{eff}}/\bar{n} \tag{3}$$

where $\overline{\delta n_{eff}} \equiv \overline{\overline{\delta n_{eff}}}(z')$ is the approximate index change, and $\bar{n}$ is a reference refractive index that is approximately equal to the average between core and cladding refractive index,

$$\overline{\beta} = \omega \overline{n} / c \tag{4}$$

is a reference wave number where $\omega$ and $c$ are the angular frequency and the speed of light, $\beta_{01}$ is the propagation constant for basic core-mode in tilted gratings, and

$$K = \pi \cos\theta / \Lambda_g \tag{5}$$

presents the transverse grating wave number, $g_f(\theta)$ and $g_b(\theta)$ present the forward and backward coupling coefficients, respectively. Just like the reference[4], $g_b(\theta) \approx g_f(\theta) \approx g(\theta)$, that is, to good approximation, the grating is equally efficient in forward and backward coupling, and for the mode s-polarized and p-polarized with respect to the gratings, similarly, to good approximation, the expressions of $g_b(s - polarized)$ and $g_b(p - polarized)$ can be equally united into $g(\theta)$ as above with an appropriate simplification.

In a similar way to obtain the reflectivity of FBG [1], the expression of reflectivity of TFBG with length $L$ can be given as follows:

$$r = \frac{\sinh^2(sL)}{\cosh^2(sL) - \dfrac{\hat{\sigma}^2}{\hat{\kappa}^2}} \tag{6}$$

Where

$$s = \sqrt{\hat{\kappa}^2 - \hat{\sigma}^2} \tag{7}$$

$$\hat{\kappa} = \overline{\beta} g(\theta) \kappa(z) \tag{8}$$

$$\hat{\sigma} = \overline{\beta} g(\theta) \sigma(z) + \beta_{01} - K \tag{9}$$

From Eqs. (3)-(9), when $\hat{\sigma} = 0$, the resonant wavelength can be found as follows:

$$\lambda_B = \frac{2\Lambda_g}{\cos\theta}\left[g(\theta)\overline{\delta n_{eff}} + n_{eff}\right] \tag{10}$$

The maximum reflectivity $r_{\max}$ is shown by

$$r_{\max} = \tanh^2\left[\frac{\pi g(\theta)\overline{\delta n_{eff}}}{\lambda_B} L\right] \tag{11}$$

For Eq. (9), when $\hat{\kappa} \geq |\hat{\sigma}|$, $s$ is a real number; otherwise, $s$ is an imaginary number, which depends on the wavelength, corresponding to the middle of central peak in the reflection spectrum from Eq. (6) and oscillating section outside the middle, respectively. Considering the null-to-null bandwidth of the spectrum, $s$ is an imaginary number, hence, we define

$$s = js' \tag{12}$$

$$s' = \sqrt{\hat{\sigma}^2 - \hat{\kappa}^2} \tag{13}$$

Substituting Eqs. (12), (13) into Eq. (6), we get

$$r' = \frac{\sin^2(s'L)}{\cos^2(s'L) - \frac{\hat{\sigma}^2}{\hat{\kappa}^2}} \tag{14}$$

When

$$s'L = \pm\pi \tag{15}$$

the first two null points appear on both sides of the central peak in the reflection spectrum, the corresponding wavelength $\lambda_0$ meets the following equation gotten by combining Eqs.(3),(4),(5),(8),(9),(13)and (15), then arranging,

$$\left(\frac{\cos^2\theta}{\Lambda_g^2} - \frac{1}{L^2}\right)\lambda^2 - \frac{4\cos\theta}{\Lambda_g}\left[n_{eff} + g(\theta)\overline{\delta n_{eff}}\right]\lambda + 4\left[n_{eff} + g(\theta)\overline{\delta n_{eff}}\right]^2 - \left[g(\theta)\overline{\delta n_{eff}}\right]^2 = 0 \tag{16}$$

If introducing new variables

$$\xi = \frac{\cos^2\theta}{\Lambda_g^2} - \frac{1}{L^2} \tag{17}$$

$$\psi = -\frac{4\cos\theta}{\Lambda_g}\left[n_{eff} + g(\theta)\overline{\delta n_{eff}}\right] \tag{18}$$

$$\zeta = 4\left[n_{eff} + g(\theta)\overline{\delta n_{eff}}\right]^2 - \left[g(\theta)\overline{\delta n_{eff}}\right]^2 \tag{19}$$

then Eq. (16) is simplified by

$$\xi\lambda^2 + \psi\lambda + \zeta = 0 \tag{20}$$

Solving the equation above, we get

$$\lambda_{0\text{-}1,2} = \frac{-\psi \pm \sqrt{\psi^2 - 4\xi\zeta}}{2\xi} \tag{21}$$

and the null-to-null bandwidth of the spectrum

$$\Delta\lambda_0 = \left|\lambda_{0\text{-}1} - \lambda_{0\text{-}2}\right| = \frac{\sqrt{\psi^2 - 4\xi\zeta}}{\xi} \tag{22}$$

## 3. Simulations and discussion

In this section, we discuss the relation between bandwidth $\Delta\lambda_0$ and tilt angle $\theta$, refractive index change $\overline{\delta n_{eff}}$ and length $L$ of tilted gratings. The fiber parameters, in accordance with reference [4], are chosen with core radius $a = 2.625um$, core index $n_{co} = 1.44792$ and cladding index $n_{cl} = 1.44$. According to the principle of optical fiber optics [6], they are single mode fibers (SMFs) with $n_{eff} = 1.44217$, approximately, thereby the grating period $\Lambda_g$ is $0.53738um$ for the wavelength of $1.55um$ at tilt angle of zero. Note that in the next discussion, we neglect the tiny changes of effective refractive index $n_{eff}$ and coupling coefficients $g(\theta)$ for the resonant wavelengths shifting towards long-wavelength side with increasing tilt angle.

Fig. 2 shows the calculated reflection spectra versus tilt angle, from which, we can see that the null-null bandwidths of resonant spectra decrease obviously with increasing tilt angle, and that the resonant wavelengths of spectra move towards the side of long-wavelength. But the maximum reflection is decreased with increasing tilt angle, specially, at the tilt angle of about $8°$, the reflection decreases sharply.

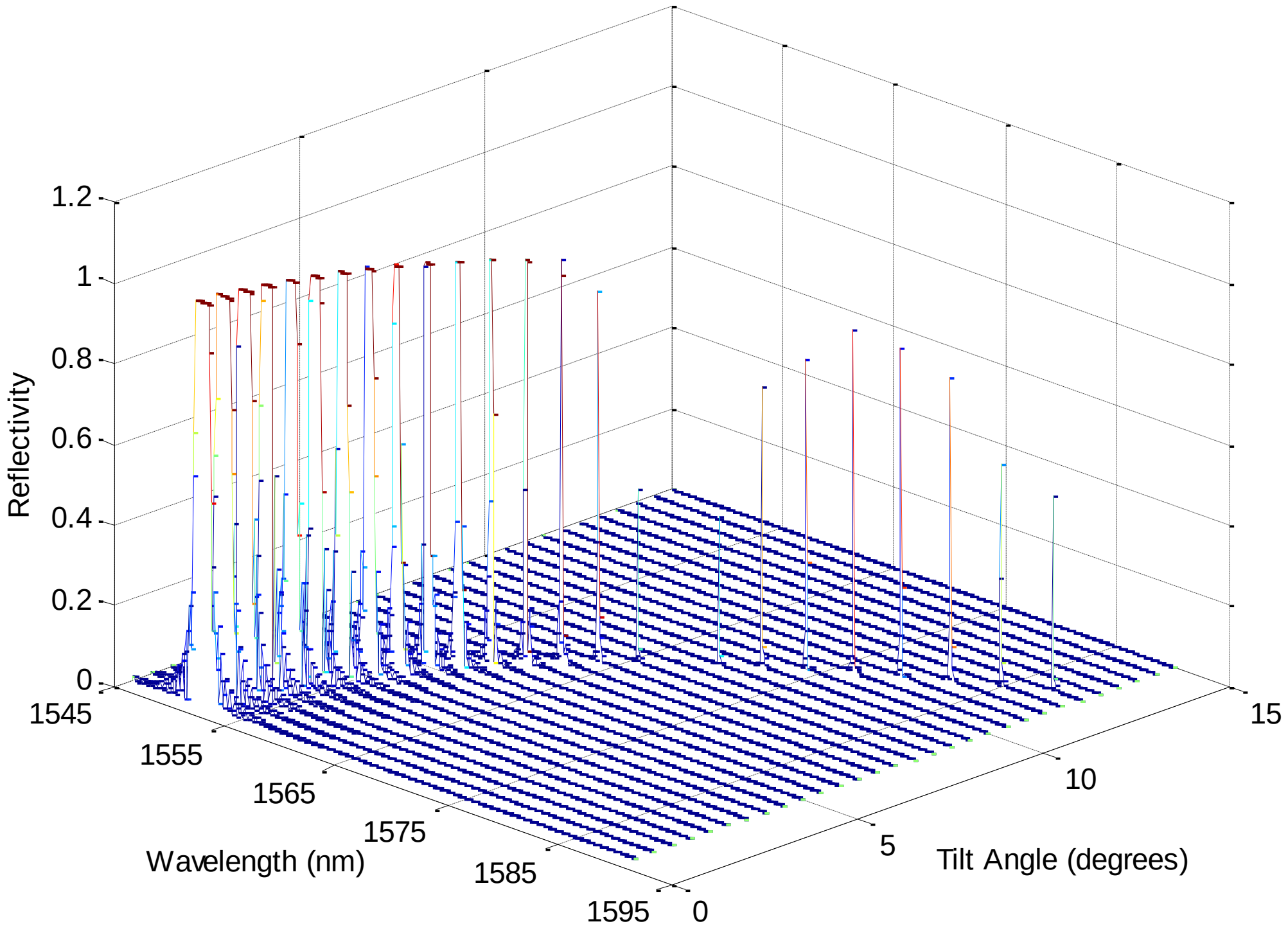

Fig. 2. Calculated tilted grating reflection spectra versus tilt angle with $\overline{\delta n_{eff}} = 2\times10^{-3}, L = 10mm$.

If we define $r_{\max} = 0.95$ that is desired maximum reflectivity, we can plot two three-dimensional diagrams in Fig. 3 and 4, describing the relation between the bandwidth, tilt angle and refractive index change, and the relation between the needed length of tilted gratings and these parameters, respectively. From the two figures, we can conclude that in the condition of reflectivity 95%, without considering the shift of central wavelength, if a very narrow bandwidth is wanted to get, the grating parameters must be obtained in the ravines in Fig. 3, and the ridges in Fig. 4, in other words, the modulation amplitude and length of grating are set nearby the tilt angles of $8°, 14°$ and $20°$. Near these angles, the grating tilt reduces the bandwidth to more narrow degree. If the index modulation amplitude is smaller in these tilt angles, the bandwidth will become narrower, however, the longer fiber gratings is needed. It is also revealed that for the tilted fiber gratings with large index modulation amplitude, the grating tilt compresses the bandwidth more fiercely, but at the expense of large grating length.

Finally, we randomly extract a point from Fig. 3 and 4, the tilt angle, effective index change and length of gratings are $8°$, $2\times10^{-3}$ and $25.82cm$, respectively. With these grating parameters, the reflection spectra is calculated in Fig. 5, from which, we can see the bandwidth and maximum reflectivity accord well with the corresponding values of the three-dimensional diagrams.

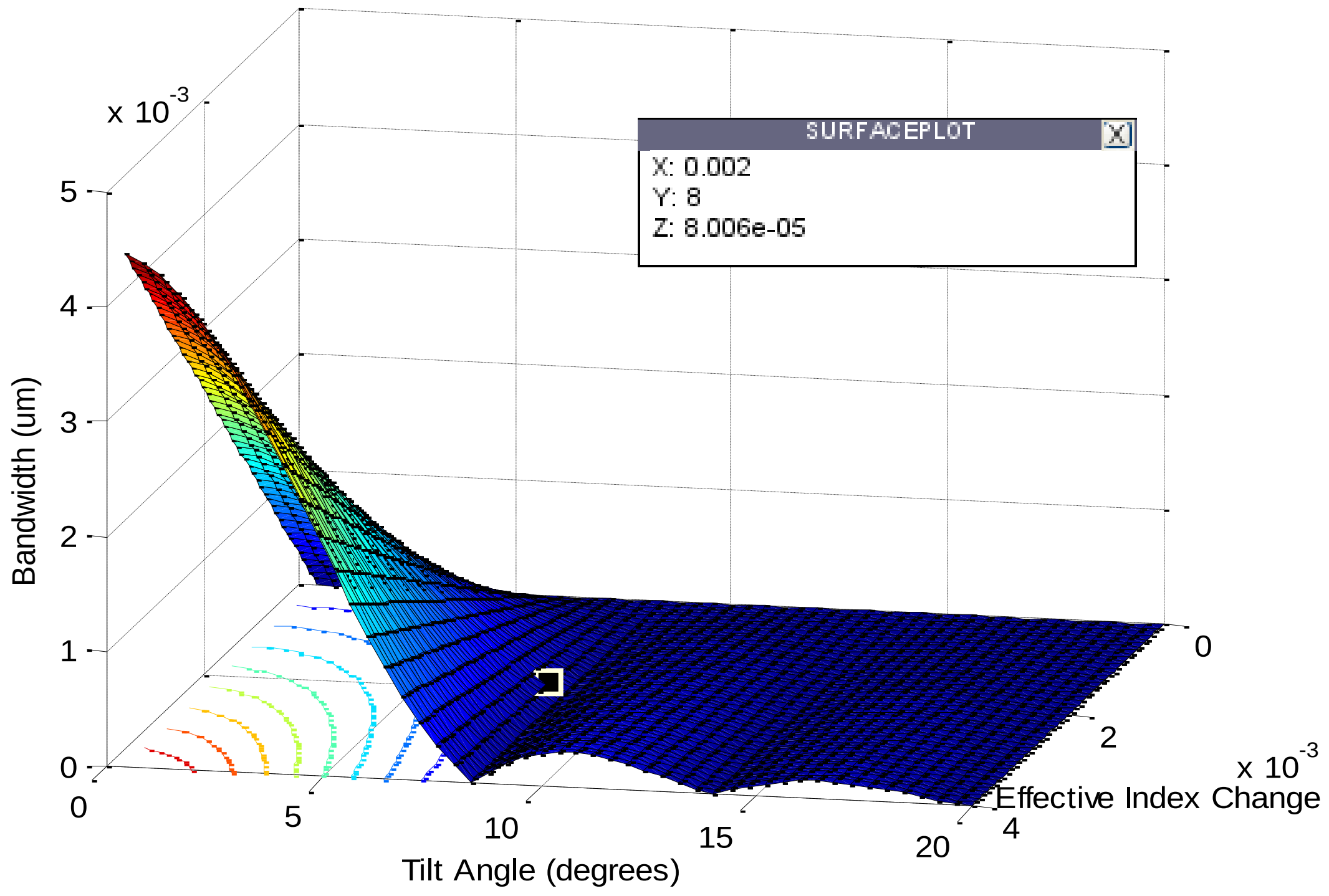

Fig. 3. Bandwidth of tilted gratings versus tilt angle and effective index change when $r_{max} = 0.95$.

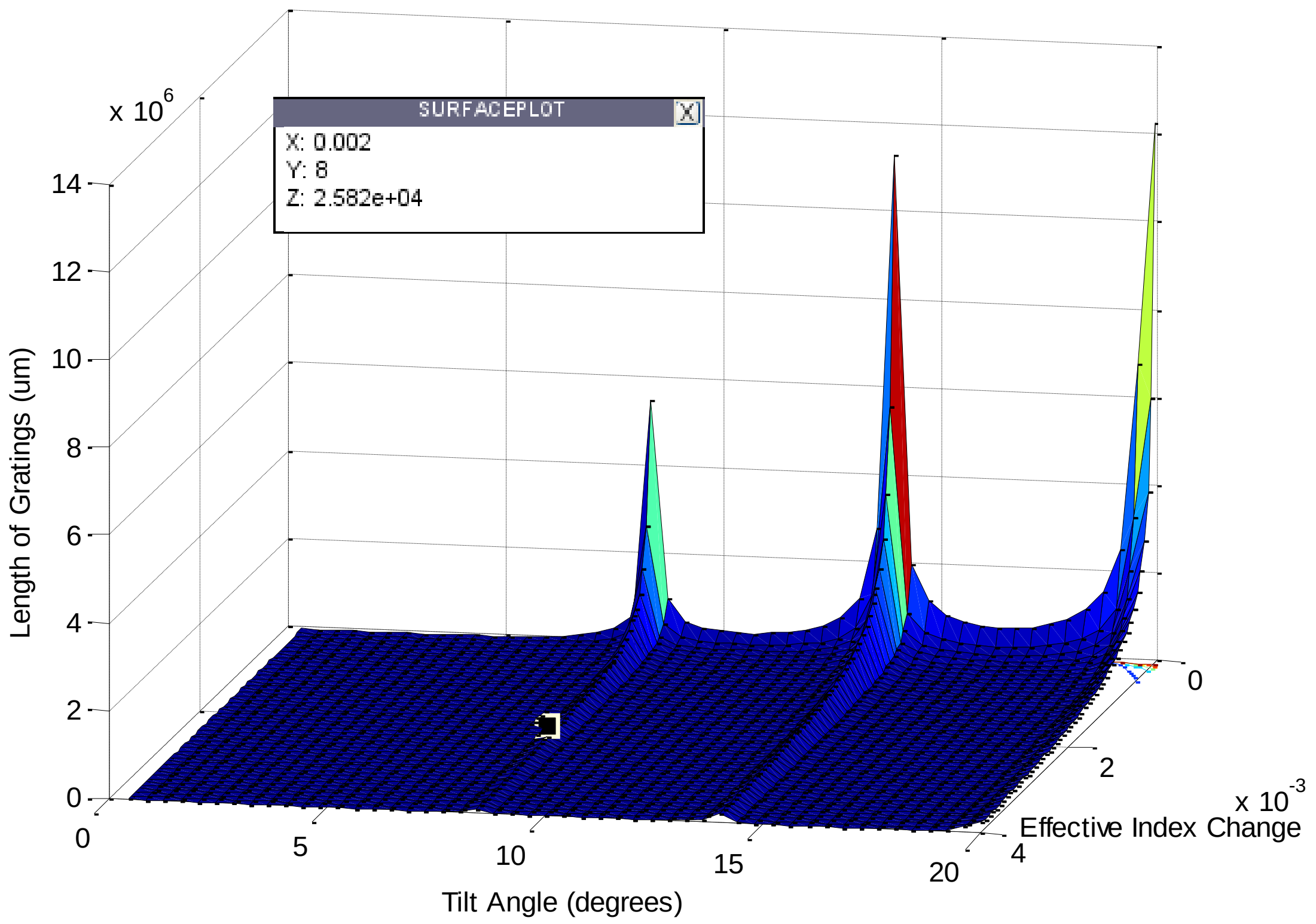

Fig. 4 Length of tilted gratings versus tilt angle and effective index change when $r_{\max} = 0.95$.

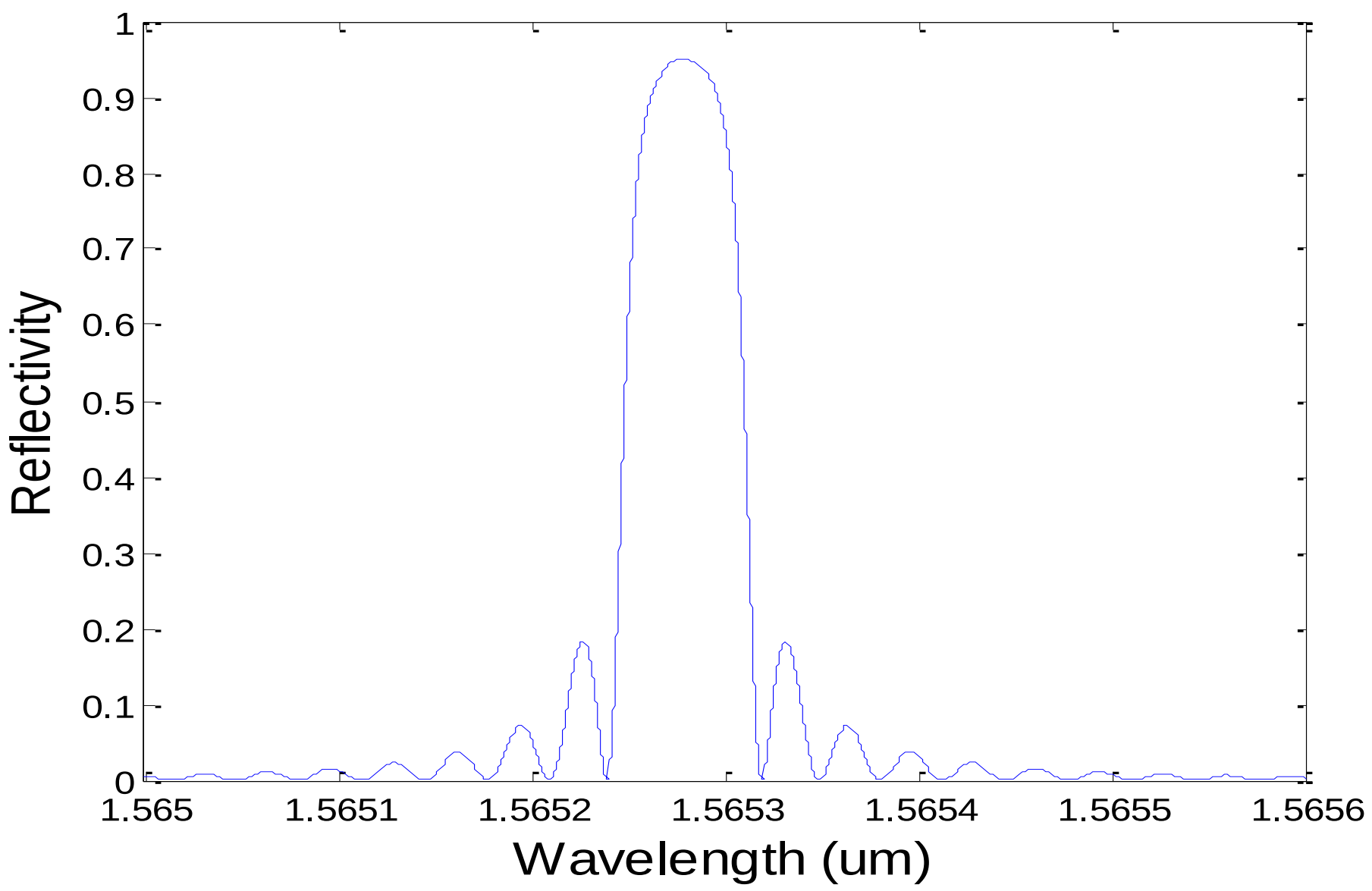

Fig. 5 Calculated reflection spectra of tilted grating with $\overline{\delta n_{eff}} = 2 \times 10^{-3}, L = 25820\,um, \theta = 8°$.

## 4. Summary

For tilted fiber gratings, on the couple-mode equations, we derive the expression of the reflectivity and resonant wavelength, explicit various parameters and variable, more importantly, deduce the formula of bandwidth originally. Numerical simulations and analyses are performed, and to make clear the comprehensive effects of grating length and modulation amplitude on bandwidth, two three-dimensional diagrams are plotted in the case of the maximum reflection of 95%. There are some special tilt angles, for instance, about $8°$, where the bandwidth is ultra-narrow, and the maximum reflectivity is still high as much as 95%. It is useful to narrow-band filter, narrow line laser and Wavelength Division Multiplexing (WDM) in optical communication and fiber sensing systems [7].

**Acknowledgement**

This work is supported by the National Basic Research Program of China (2013CB707500), the Innovation Fund Project for Graduate Student of Shanghai (JWCXSL1302), and partly supported by the Shanghai Leading Academic Discipline Project (No.S30502).

**References**

[1] T. Erdogan, Fiber Grating Spectra, J. Lightwave Technol. 15 (1997) 1277-1294.

[2] K. O. Hill, G. Meltz, Fiber Bragg grating technology fundamentals and Overview, J. Lightwave Technol. 15(1997) 1263-1276.

[3] S.A. Vasil'ev, O.I. Medvedkov, I.G Korolev, A.S Bozhkov, A.S. Kurkov and E.M. Dianov, Fibre gratings and their applications, IEEE J. Quantum Electron. 35 (2005) 1085–1103.

[4] T. Erdogan, J. Sipe, Tilted fiber phase gratings J. Opt. Soc. Am. A , 13(1996) 296-313.

[5] X. Ou, L. Shaohua, L. Yan, L. Bin, D. Xiaowei, P. Li, J. Shuisheng, Analysis of spectral characteristics for reflective tilted fiber gratings of uniform periods, Optics Communications 281(2008) 3990-3995.

[6] C. Tsao, Optical Fibre Waveguide Analysis. (Oxford University Press, Walton Street, 1992), Part 3.

[7] . E. Gerd, Keiser, A review of WDM technology and applications Optical Fiber Technology, 5(1999) 3-39.